\documentclass[10pt,aps,prl,showpacs,superscriptaddress,twocolumn,citeautoscript]{revtex4-2}

\usepackage{graphicx}  
\usepackage{dcolumn}   
\usepackage{bm}        
\usepackage{amssymb}  
\usepackage{amsmath}
\usepackage{hyperref,color,braket}
\usepackage{epsfig,wrapfig}
\usepackage{gensymb}
\usepackage{tikz}

\newcommand{\beq}{\begin{equation}}
\newcommand{\eeq}{\end{equation}}

\begin{document}
\title{Observation of Two-Dimensional Acoustic Bound States in the Continuum}
\author{Marc Mart\'i-Sabat\'e}
\affiliation{GROC, UJI, Institut de Noves Tecnologies de la Imatge (INIT), Universitat Jaume I, 12071, Castell\'o de la Plana, Spain}
\author{Junfei Li}
\affiliation{Department of Electrical and Computer Engineering, Duke University, Durham, North Carolina 27708, USA}
\author{Bahram Djafari-Rouhani}
\affiliation{IEMN, University of Lille, Cité scientifique, 59650 Villeneuve d’Ascq, France}
\author{Steven A. Cummer}
\affiliation{Department of Electrical and Computer Engineering, Duke University, Durham, North Carolina 27708, USA}
\author{Dani Torrent}
\email{dtorrent@uji.es}
\affiliation{GROC, UJI, Institut de Noves Tecnologies de la Imatge (INIT), Universitat Jaume I, 12071, Castell\'o de la Plana, Spain}
\date{\today}

\begin{abstract}
The design of devices based on acoustic or optical fields requires the fabrication of cavities and structures capable of efficiently trapping these waves. A special type of cavity can be designed to support resonances with a theoretically infinite quality factor, named bound states in the continuum or BICs. The experimental measurement of such modes is still a challenging problem, as they are, by definition, not accessible from external  perturbations. Therefore, current reported works rely on indirect measurements that are based on the traces left by these modes on external properties of one-dimensional systems. Here we report on the theoretical design and experimental realization of a two-dimensional, fully open acoustic resonator supporting BICs. This BIC, whose symmetry is chosen during design by properly tailoring the geometrical properties of the system, is completely accessible and allows for the direct measurement of the whole pressure field and properties. We experimentally demonstrate its existence with high quality factor and field enhancement properties.    
\end{abstract}
\maketitle

\section{Introduction}


Controlling the propagation and localization of waves is of paramount importance for a large number of modern applications based on the use of the energy or information carried out by these waves, including optical communications, quantum computing and sensing and imaging. One of the most efficient methods to achieve this control are the so-called embedded eigenstates, or bound states in the continuum (BICs), which have attracted great interest in recent years due to their many advantageous properties. BICs are modes in a system whose energy belongs to the radiation part of the spectrum while remaining spatially confined with an infinite lifetime. These modes were theoretically predicted just after the emergence of quantum mechanics by von Neumann and Wigner \cite{neumann1929merkwurdige}. Since then, BICs have been designed and analyzed, resulting in different kinds and classifications for them, not only in quantum physics but also in photonics \cite{molina2012surface,miroshnichenko2010fano,hsu2016bound,hsu2013bloch,sadreev2021interference,bulgakov2008bound,zhen2014topological} and acoustics \cite{friedrich1985interfering,huang2022general,quotane2018trapped,jin2017tunable,mizuno2019fano,PhysRevB.106.085404}. 
Their infinite quality factor makes them promising components of filters and resonators for classical waves, and for enhancing wave-matter interaction \cite{kodigala2017lasing,koshelev2019meta,wu2020room,koshelev2020subwavelength}.
Their existence has been proved experimentally both in photonics \cite{gansch2016measurement,capasso1992observation,PhysRevLett.107.183901,shi2022planar,chen2023observation} 
and acoustics \cite{cobelli2009experimental,huang2020extreme,huang2022topological,huang2021sound,amrani2021experimental}. 

Concerning acoustics, experimental measurements were first performed with a BIC created by a closed cavity attached to a one-dimensional waveguide by a small port \cite{huang2021sound}. This zero-dimensional BIC is enclosed within the cavity, and its quality factor was estimated through indirect measurements. Another experimental measurement was done using an open resonator embedded in a one-dimensional waveguide. \cite{PhysRevApplied.19.054001} The quality factor was again estimated using the waveguide's transmission coefficient. Building upon this foundation, another experimental effort \cite{huang2022topological} employed two separated cavities connected to a one-dimensional waveguide, in which the interaction between them creates a one-dimensional BIC (a Friedrich-Wintgen BIC). 

However, the acoustic pressure field in this 1D case remains inaccessible to experimental measurements, and in both the previous 0D and 1D demonstrations, key properties are once again only indirectly deduced from the analysis of the reflection spectrum. By definition, BICs are confined and isolated from the rest of the system, implying that their excitation and measurement is not possible without altering the geometry by opening an input/output channel and thus leaking significant energy. Measuring directly the fields of a BIC mode is crucial to exploit the main characteristics of these systems; not only by their extreme confinement and its divergent quality factor, but also to probe the extreme field enhancement around their resonant frequency. This kind of measurement will open the field to real applications in which the properties of BICs will be actively exploited. Moreover, these previous efforts have been limited to zero- and one-dimensional acoustic BICs that do not exhibit the largely open boundaries that give BICs their most remarkable properties.

In this work we present the design of a two-dimensional fully open resonator that supports BICs, and we fully characterize the BIC properties for acoustic waves by direct measurement. The system is formed by a set of precisely designed blind holes arranged as a regular polygon in a two-dimensional acoustic waveguide. By measuring the spatial distribution of the acoustic pressure field with fine frequency resolution, we obtain a complete picture of the BIC resonance, confirming good agreement between the designed and realized properties. Our basic BIC design is suitable for the generation of many multipolar modes, that might be foreseen useful for a broad range of applications based on the control of classical and quantum waves.

\begin{figure*}
\centering
\epsfig{file=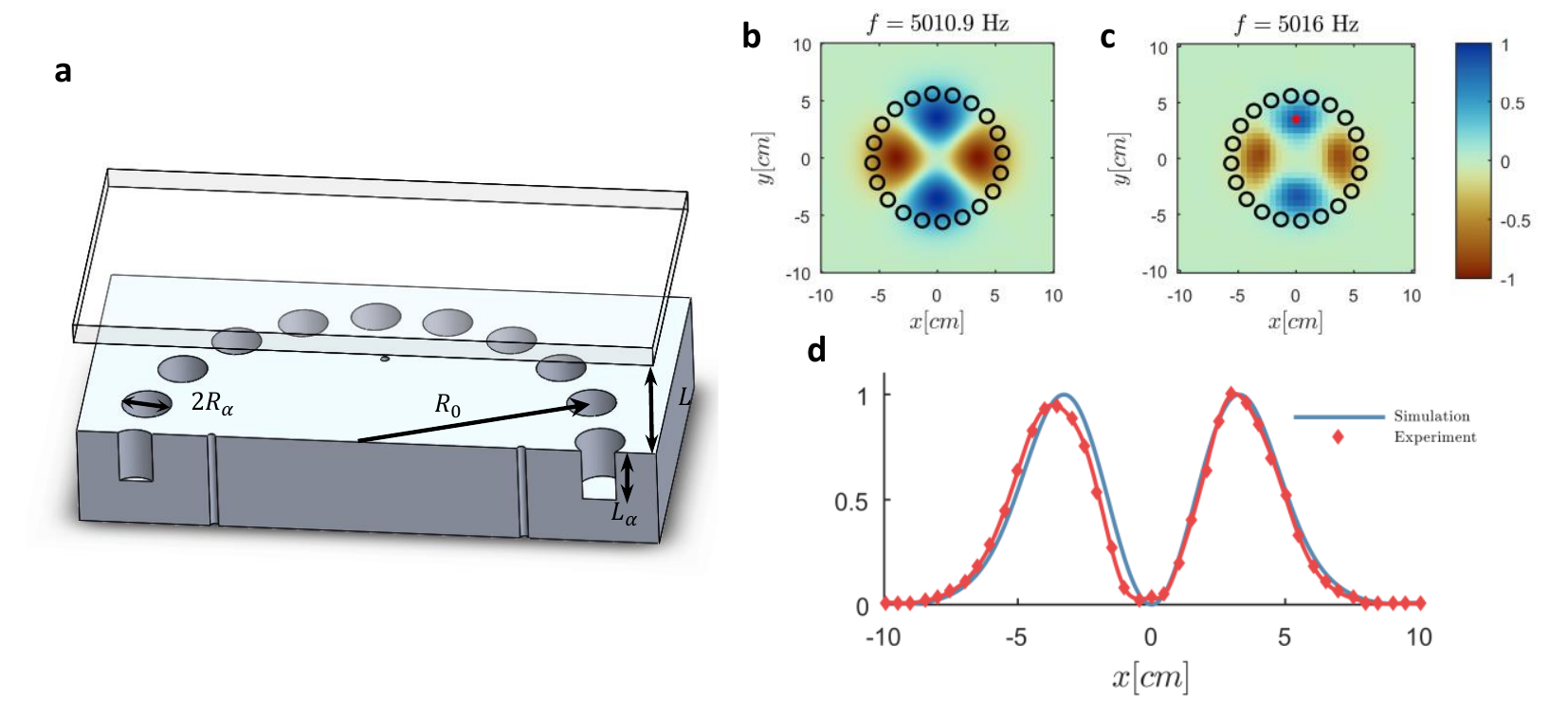, width = \textwidth}
\caption{\label{Fig:Figure_1_montage}{Designed bound state in the continuum and achieved performance. Panel \textbf{a} shows an illustration of half of the designed plate, with its glass cover and the input channels. Panels \textbf{b} and \textbf{c} show the BIC normalized real pressure field both in simulation (\textbf{b}) and in experiment (\textbf{c}). Panel \textbf{d} depicts the normalized absolute pressure field for the line $y = 0$, showing good agreement between simulation and experiment.}}
\end{figure*}


\section{Results}
\label{sec:Results}

Figure \ref{Fig:Figure_1_montage} summarizes the main result obtained in this work. Panel \textbf{a} shows the designed structure supporting a BIC, which consists of a cluster of $N$ identical blind holes of radius $R_\alpha$ and depth $L_{\alpha}$ drilled on the top surface of an acoustically rigid plate and placed regularly along the perimeter of a circumference of radius $R_0$. A second plate is placed at a distance $L$ from the top surface of the bottom plate, forming therefore a two-dimensional waveguide for acoustic waves. Only half of the structure is shown in the figure, so as to appreciate the depth of the holes. Previous works have explored the presence of high quality modes in this geometrical configuration for elastic\cite{movchan2019platonic,putley2021whispering,PhysRevResearch.5.013131,app11104462} and electromagnetic waves \cite{bulgakov2018nearly,bulgakov2018fibers,kuhner2022radial}, relating them to BICs. Panels \textbf{b} and \textbf{c} depict the theoretical prediction and the experimental characterization, respectively, of the mode analyzed in this work. The numerical prediction (\textbf{b}), simulated in COMSOL's Acoustic module frequency domain is obtained by sweeping in frequency and keeping the distribution with the biggest amplitude. The experimental field shown in panel \textbf{c} depicts the pressure field distribution at the resonance frequency. The experimentally measured quality factor for this resonance is $Q = 182$. Panel \textbf{d} shows a cross section of panels \textbf{b} and \textbf{c} along the line $y = 0$. A good agreement is found between simulation and experimental measurements. 
Next section will show the design process of the geometrical parameters of both the holes and the plate to obtain a resonant mode with an infinite quality factor (purely real eigenfrequency), obtaining then a BIC mode. Since we have employed an analytical method with some approximations, finite element simulations using a commercial software  (COMSOL) have been done, fine tuning the parameters of the system and studying the behavior of the mode under a more realistic environment. Finally, real experimental measurements have been conducted, and the experimental results will be shown and discussed. 

\textbf{2D BIC Design.} 
The eigenfrequencies of the system shown in figure \ref{Fig:Figure_1_montage} can be found by the mode-matching method\cite{torrent2018acoustic,PhysRevLett.108.174301}  as explained in detail in the supplementary material. This method applies boundary conditions (continuity of pressure and normal velocity field) in the region of contact of two different domains (waveguide and cavities in our case) and simplify the system of equations by projecting the modes into another base of orthogonal modes. The quality factor of an eigenmode is inversely proportional to the imaginary part of its eigenfrequency, thus we can define a BIC in acoustics as those modes having zero imaginary part in its eigenfrequency. For a cluster of holes drilled in an acoustically rigid cavity, a real eigenfrequency can be found as long as the following equation is satisfied
\beq
   \cot{(k_bL_\alpha)} + 2N(-1)^{\ell}I_{\ell}(k_b)=0,
    \label{eq:BICCondition}
\eeq
with $I_{\ell}$ being
\beq
    I_{\ell}(k_b) = \int_0^{+\infty}\frac{k_b}{q_k k}\cot{(q_kL)}J_\ell^2(kR_0)J_1^2(kR_\alpha)dk,
    \label{eq:IntegralEquationMainArticle}
\eeq
and $R_\alpha, L_\alpha, R_0$ and $L$ being the radius and depth of the holes, the radius of the cluster and the height of the waveguide, respectively. The integer number $\ell$ is a label that defines the multipolar order of the mode (see equation ($15$) in the supplementary material for further details), $k_b = \omega/c_0$, $c_0$ is the speed of sound, and it has been set $c_0 = 344 m/s$ and $q_k = \sqrt{k_b^2 - k^2}$. 

Equation \eqref{eq:BICCondition} is a transcendental equation for the eigenfrequency $\omega=c_bk_b$, consequently it is not efficient to select the geometry and then try to find the frequency at which the BIC is obtained. Instead, we can select the frequency at which we wish the BIC, then fix $L$, $R_0$ and $R_\alpha$ and, after the calculation of $I_{\ell}(k_b)$, we can use \eqref{eq:BICCondition} to obtain $L_\alpha$. As can be seen, $I_\ell (k_b)$ is always real, and the left hand side of equation \eqref{eq:BICCondition}  can take any value in the real axis, consequently there will be always a right cluster dimension that corresponds to a BIC at any given frequency and for different multipoles for various values of $\ell$.

The number of degrees of freedom, including the multipolar order of the mode, for this design is very large, however, if we want to obtain reasonable dimensions for the cluster that fits our experimental constraints, a systematic design of the cluster has to be done. 

The approach that we followed allowed us to have a considerable control over the dimensions of the cluster. We begin by selecting the frequency at which we want the BIC to be found, which is $f_0=5$ kHz for experimental convenience. Next, selecting a symmetry of the field of $\ell=2$, we set the radius $R_0$ of the cluster such that $k_bR_0$ is the argument of the first zero of the second order Bessel function, $j_{2,1}=5.1356$. The reason for this choice is that it will minimize the field at the border of the cluster, so that we can expect that the quantity $I_{\ell}(k_b)$ will be small and then $k_bL_\alpha$ will be close to $\pi/2$, giving a reasonable size for the length $L_\alpha$. Finally, we get $R_0=5.61 cm$. 

Once the radius of the cluster has been selected, the radius of each hole can be established just by taking into account some constraints. The first constraint is about the maximum size of the radius of the hole. The distance between two adjacent holes in a circular cluster is $2R_0\sin{(\pi/N)}$ \cite{PhysRevResearch.5.013131}; therefore, the radius is restricted to $R_\alpha < R_0\sin{(\pi/N)}$ to avoid overlapping of the holes. Furthermore, if the radius of the holes is too large, the theoretical model developed here might fail, since we have used the mono-mode approximation in our equations, as explained in the supplementary material. The second constraint involves the minimum size of the radius. If it is too small, in practice we will have very narrow channels for which the sound wave will be strongly dissipated, and the experimentally observed mode will have a poor quality. We have chosen $R_\alpha = 2/3R_0\sin{(\pi/N)}$, so that the desired effect is produced. Selecting $N=20$ we get $R_\alpha= 5.9$ mm. Finally, we select $L=2.5cm$ for practical reasons and equation \eqref{eq:BICCondition} gives $L_\alpha= 1.36$ cm. Since the mode-matching method is not taking into account some evanescent fields near the surface of the holes, we choose to use the COMSOL simulations to compensate for it and find the exact solution for the BIC.

Equation \eqref{eq:BICCondition} also shows that, for a given geometry of the cluster, changing the height of the waveguide will change the frequency of the mode. It also includes the uncovered scenario, in which we remove the top of the waveguide and have an open system. In this case, no BIC condition can be achieved, as explained in the suplementary material, however we also characterized experimentally this mode to measure its quality factor. Numerical simulations and experimental results can be found in detail in the supplementary material, proving the existence of the resonance. For the covered case, after fine tuning the height of the plate ($L$), numerical simulations show a good agreement with the analytical design, as can be seen in figure \ref{Fig:Figure_1_montage} panel \textbf{b}. As expected, the field is contained in the inner part of the circle of holes, and the symmetry is the one given by the selected multipolar index $\ell$.

\textbf{Experimental measurements.} Measurements were carried out to characterize the design experimentally and confirm the existence of a BIC, although in practice dissipation is always present and we have opened an inlet/outlet channel from the near field, which makes the mode present a finite quality factor. Experimental results show good agreement with predicted and simulated features. In figure \ref{Fig:Figure_1_montage}, panels \textbf{b}, \textbf{c} and \textbf{d} prove that the scattering field at $5016$ Hz looks the same as the one found in simulations, that is to say, to the designed BIC distribution. This excited field is completely confined inside the circle of radius $R_0$; the leakage of energy to the outside of the circle is evanescent, which means that these waves outside the circle do not carry energy to the far field. The experimental field shown in panel \textbf{c} is obtained from the real part of the Fourier transform of the field excited by means of a gaussian pulse, selecting the Fourier component which presents the highest amplitude.

\begin{figure*}
\centering
\epsfig{file=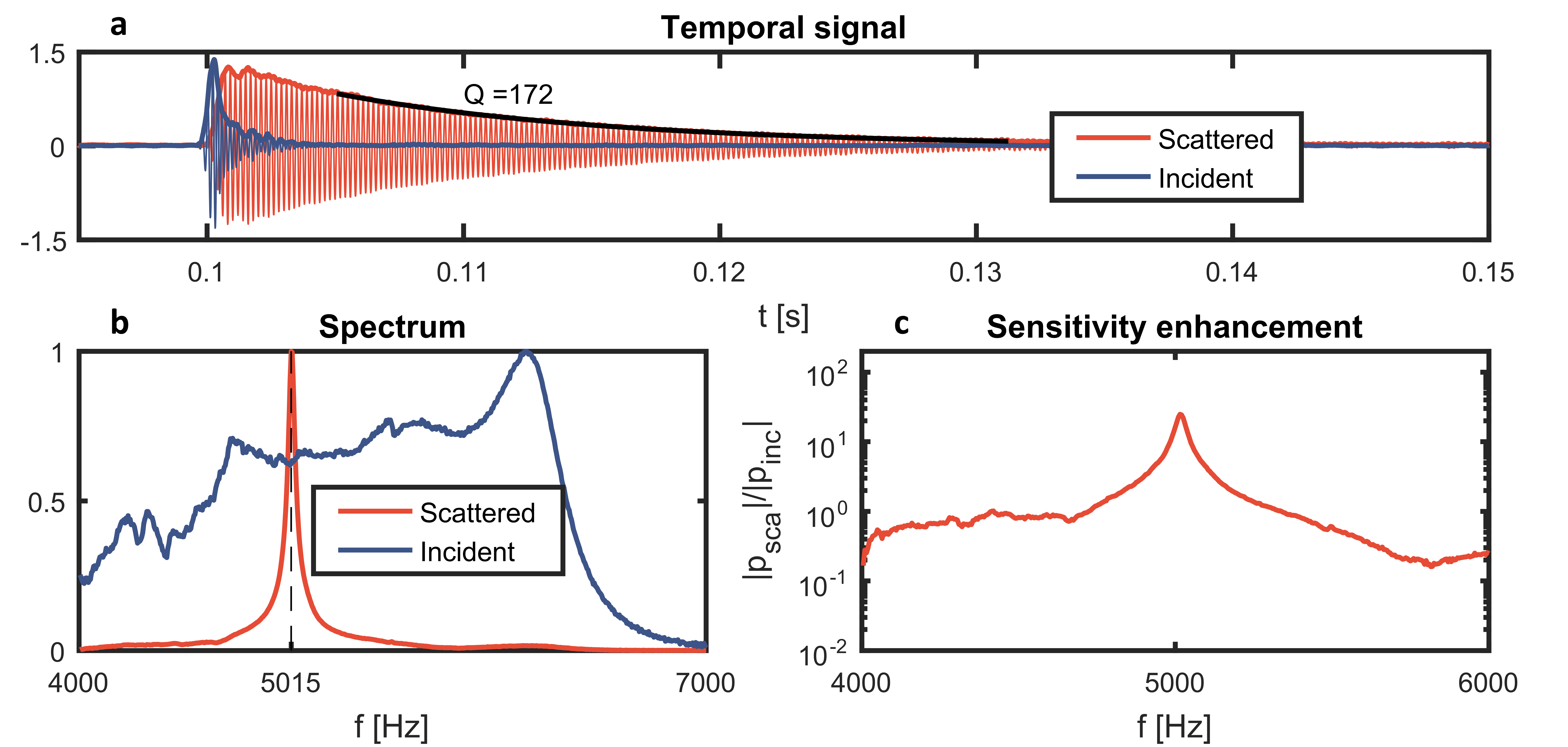, width = \textwidth}
\caption{\label{Fig:Figure_3_montage}{Experimental results obtained for the $\ell = 2$ designed plate. Panel \textbf{a} shows the incident and the scattered signal and envelope at a given point ($x = 0$, $y = 35$ mm, $z = 22$ mm). The black line is the fitting by a decaying exponential used to estimate the quality factor of the resonance. Panel \textbf{b}} shows the spectra of both signals in normalized units. Panel \textbf{c} shows the ratio between the scattered spectrum and the input spectrum. A peak at $5015$ Hz is seen, showing an energy enhancement at this frequency due to the presence of the cluster of holes.}
\end{figure*}

Four speakers are placed under the aluminum plate oriented upwards and, on top of them, four $2$ mm diameter passing holes connect the top and bottom surface of the plate, letting the energy from the speakers enter the cavity. The input signal to the speakers is a gaussian pulse centered at $5$ kHz and spanning from $4$ kHz to $6$ kHz. The incident field is measured by covering up the holes, thus leaving a flat, empty waveguide. As can be seen in figure \ref{Fig:Figure_3_montage} panel \textbf{b}, the incident spectrum at the source point ($x = 0$, $y = 35$ mm, $z = 22$ mm, indicated in figure \ref{Fig:Figure_1_montage} panel \textbf{c} by a red dot) is different from a Gaussian shape, due to the non-uniform frequency response in the acoustic source, introduced by the holes in the plate, the speakers and the cavities behind them.

Nevertheless, when the cluster of holes is considered, the scattered field presents a sharp resonance at $5016$ Hz. Figure \ref{Fig:Figure_3_montage} panel \textbf{b} shows the two normalized spectra. The spectrum of the excited field (blue line) shows that other frequencies different from $5$ kHz are generated, but their relative amplitude is much lower than the resonance. The temporal signal in panel \textbf{a} also agrees with this interpretation. The duration of the incident pulse is less than $5$ ms at the measurement position, while the scattered field rings at least $50$ ms. The temporal envelope, together with the narrow spectral content, states that the mode at $5$ kHz is excited and retained in the inner part of the cluster for a long time, as can be seen in the video of the supplementary material. Such a long ringing time allows us to directly measure the quality factor, instead of calculating FWHM. The black curve in figure \ref{Fig:Figure_3_montage} panel \textbf{a} shows the fitting of the curve by a decaying exponential ($x(t) = A·e^{-\omega_i t}$). The quality factor is then estimated as $Q = \omega_0/2\omega_i$, where $\omega_0$ is the resonant frequency and $\omega_i$ is the imaginary frequency term. In the example shown in the figure, the measured $Q$ is $172$. The quality factor should be independent of the position at which the measurement is done; in fact, we have measured many other points to confirm that the estimation of the quality factor is consistent.

Figure \ref{Fig:Figure_3_montage} panel \textbf{c} shows the ratio between the scattered and incident spectrum. There is a notably sharp peak around the $5$ kHz, signifying a substantial sensitivity enhancement, approximately twenty-five times larger. Such a feature makes our BIC structure particularly appealing for sensing applications. The measured acoustic field at peak frequency, shown in Fig \ref{Fig:Figure_1_montage} panel \textbf{c}, demonstrates almost perfect agreement with the theoretical prediction.


%
%

\begin{figure}
	\epsfig{file=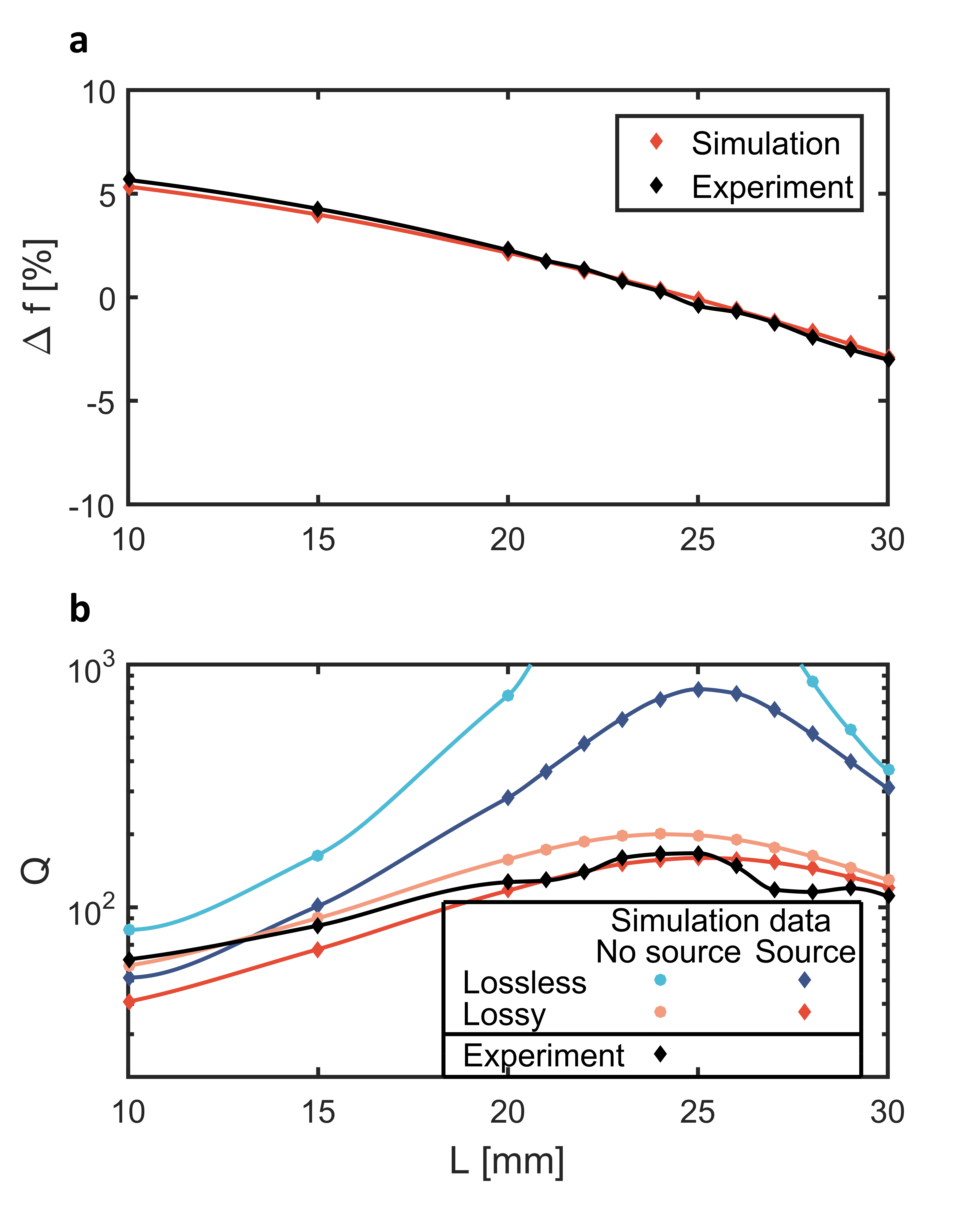, width = 0.45\textwidth}
	\caption{\label{Fig:fQEvolution}{\textbf{a} Dependence of the resonant frequency of the mode with the height $L$ of the cavity keeping all the other geometrical parameters fixed. \textbf{b} Quality factor of the resonance as a function of the covered plate distance. Black rhombus represent the experimental results, while the rest of data is simulation, considering lossy or lossless material and considering the presence of the input channel in the system. As can be seen, the quality factor of the BIC is far from being infinite, and the two loss mechanisms (lossy material and input channel) explain the behavior of the experimental results.}}
\end{figure}

A remarkable property of the system studied in this work is its easy reconfigurability, since the BIC condition strongly depends on the geometrical parameters of the waveguide and the cluster, which means that the quality factor can be easily controlled by adjusting the geometry, such as the height of the waveguide ($L$). Figure \ref{Fig:fQEvolution} panel \textbf{a} shows the evolution of the resonant frequency of the mode with the height $L$ of the cavity. As we see, the relative variation of the eigenfrequency is very small, although the quality factor strongly depends on this parameter, as explained below.

Figure \ref{Fig:fQEvolution} panel \textbf{b} shows the evolution of the quality factor as a function of the height of the waveguide $L$ for the $\ell = 2$ mode. The experimental results are depicted in black. The quality factor is not divergent as we approach to the theoretically predicted BIC, due to the presence of dissipation in the real system. Nevertheless, it is shown that these results match the ones found in simulation considering losses in the material and in the excitation system. Blue dots represent the simulation values obtained for eigenfrequency simulations considering no losses in the material and no losses due to the excitation system. It is seen that, in the interval from $L = 21$ mm to $L = 26$ mm, the quality factor is higher than $1000$. The range in the figure has been limited so as to appreciate the shape of the curves considering noise. However, a real eigenvalue (and thus, an infinite quality factor) can be found for $L = 24$ mm. Blue diamonds in figure \ref{Fig:fQEvolution} panel \textbf{b} represent eigenfrequency simulations considering no losses in the material, but losses in the excitation system (the losses have been introduced in simulation by considering cylindrical wave radiation boundary condition at the bottom surface of the $4$ input channels). The results show that the quality factor cannot reach an infinite value as in the previous case; the quality factor will always remain under $1000$. The maximum is now found for $L = 25 mm$ ($Q = 792$). This result agrees with what it is known about BICs; they are states whose energy remains confined for an infinite lifetime, without leakage into the bulk. The system we are working with is reciprocal, meaning that, neither the energy can be leaked to the outside, nor the energy coming from the outside will excite the BIC. Thus, the mode is completely isolated from the outside. In order to achieve the excitation of the state, an input/output channel must be created. In the case of this experiment, this is the role of the four holes located in the interior of the circle and connecting the top and bottom surface of the aluminum plate, where the speakers are placed. The energy is then able to flow from the speakers into the waveguide, where the BIC mode will be excited. Consequently, these channels will allow small energy leakage from the inside the cluster to the outside. 

Red dots in figure \ref{Fig:fQEvolution} panel \textbf{b} represent eigenfrequency simulations with material loss (adding an imaginary component to the speed of sound) but without considering leakage from the excitation system. The quality factor threshold is even lower than in the previous case, indicating that this loss mechanism is more crucial than the previous one, even if the percentage of losses in air has been estimated to be very low ($0.25\% c_0 = 0.86 m/s$). Finally, red diamonds in figure \ref{Fig:fQEvolution} panel \textbf{b} represent eigenfrequency simulations with both loss mechanisms. Experimental results match these last simulations. They have been measured under the following environmental conditions: $T = 21.5 \degree C$ and $ HR = 10\%$ (humidity rate). According to \cite{harris1966absorption}, decreasing the temperature or increasing the humidity could further reduce the inherent loss in air. By using a humidifier near the waveguide, we have measured $Q = 182$ at $T = 21.5 \degree C$ and $ HR = 40\%$. 

\section{Conclusions}
\label{sec:Discussion}

In summary, we have designed and experimentally measured an acoustic two-dimensional open resonator supporting the existence of a family of bound states in the continuum (BICs) in circular arrays of scatterers. Its performance depends on the geometrical parameters of the configuration, and can be easily tuned numerically. Our approach allows us to confine the acoustic field in a fully open space in a two-dimensional waveguide instead of inside a closed cavity, and it is robust enough to allow for the direct measurement of the field inside the waveguide without destroying the resonance. Experimental measurements agree with simulations; the pressure field is contained by the structure and there is no leakage of energy to the waveguide. Furthermore, the existence of the BIC is linked to the field enhancement at the mode frequency. Results show that the energy density is increased by more than two orders of magnitude at the BIC frequency.
The maximum measured quality factor is $182$; this result matches with what is found in simulation and it is consistent with the loss mechanisms that exist in the system, which is dominated by intrinsic acoustic losses in air. 

The properties of our design may find applications in enhanced acoustic emissions, and might be suitable for the design of acoustic filters and sensors. Also, the results shown here were previously theoretically demonstrated for elastic and optical waves, consequently a properly designed system could be used for the simultaneous control of different wave fields or even to enhance the interaction between them.

\section{Methods}

\textbf{Numerical simulations.} The full wave simulations based on finite element analysis are performed using COMSOL Multiphysics Pressure Acoustics module. The simulated domain is a cylinder with radius $R_{cyl} = 3R_0$. For the uncovered plate scenario, perfectly matched layers are adopted in the top boundary to reduce reflections, while plane wave radiation boundary conditions are applied on the sides to simulate the propagation to an infinite system. In the covered case, the perfectly matched layer is substituted by a rigid boundary condition at the height of the top plane. Speed of sound is 344 m/s (after measurement in the laboratory). Losses have been simulated by adding an imaginary component to the speed of sound. Furthermore, losses in the input channel have been simulated by applying cylindrical wave radiation boundary conditions in the bottom surface of the throughout holes connecting the bottom and top surface of the bottom plate. All boundaries between air and solid (glass or aluminium) have been considered rigid.
\newline

\textbf{Experimental apparatus.} The sample was fabricated by drilling twenty blind holes in aluminum with a CNC machine. The size of the plate is $18$x$18$x$1$ inch. The diameter of the holes is $1/2$ inch. The four throughout holes corresponding to the input energy channels have also been drilled using a CNC machine. In this case, the diameter is $5/64$ inch ($\sim 2$ mm). A $1$-inch speaker is attached to the end of each through hole, acting as the sound source.  Four speakers 
have been used as excitation system, emitting a pulse centered at $5$ kHz and spanning from $4$ kHz to $6$ kHz. The amplitude of each speaker has been normalized to obtain the same spectral response on top of the speaker. The pressure field was measured by scanning the surface with a MEMS microphone (SparkFun ADMP401) attached to a magnet following the movement of the scanning stage outside the waveguide. The signal is then collected by a DAQ (National Instruments PCI 6251). The overall scanned area is $10$ cm by $10$ cm with a step of $0.5$ cm. Each position is repeated ten times and the resulting signal is time-averaged in order to reduce noise. Measurements shown here have been done under laboratory environmental conditions ($T = 21.5^\circ C$ and $HR = 10\%$). The edge of the structure is surrounded by acoustic foam to reduce reflection, and the detected signals are time-gated to eliminate reflection. The scattered field is measured as the difference between the total field (with the cluster), and incident field (without the cluster).

\section{Acknowledgements}
This work was supported by DYNAMO project (101046489), funded by the European Union. Views and opinions expressed are however those of the authors only and do not necessarily reflect those of the European Union or European Innovation Council. Neither the European Union nor the granting authority can be held responsible for them. Marc Mart\'i-Sabat\'e acknowledges financial support through the FPU program under grant number FPU18/02725. This publication is part of the project PID2021-124814NB-C22, funded by MCIN/AEI/10.13039/501100011033/ "FEDER A way of making Europe".



%

\end{document}